\documentclass[a4paper]{jpconf}
\usepackage{graphicx}
\begin{document}
\title{Measuring Dark Matter Distribution in Directional Direct Detection}

\author{Keiko I. Nagao\footnote{This work is based on a collaboration with Tatsuhiro Naka (Nagoya University) and Mihoko M. Nojiri (KEK and Kavli IPMU).}}


\ead{nagao@sci.niihama-nct.ac.jp}
\address{Niihama National College of Technology, Department of Engineering Science\\
7-1 Yagumo  Niihama  Ehime 792-8580  Japan}

\begin{abstract}
Direct detection of dark matter with directional sensitivity offers not only measurement of  both recoil energy and direction of dark matter, but also a way to understand dark matter distribution in the Galaxy. Maxwell distribution is usually supposed as the distribution near the Earth, however, deviation from that, caused by tidal streams in the Galaxy, has been suggested. 
We explore  the possibility of distinguishing the distribution by direct detection using nuclear emulsions. 

\end{abstract}

\section{Introduction}

Non-baryonic dark matter is a hot subject for both particle physics and astrophysics. Especially, it can be a valuable clue to the physics beyond the standard model in particle physics. Many theoretical model which has a candidate for dark matter particle, such as the supersymmetric models and  models with extra dimensions, have been proposed. 

Direct detection of dark matter imposes severe bounds to such theories, as well as the relic abundance and the indirect detection. 
It gives lower bound for the interaction cross section between dark matter and nucleon. 
Event rate $R$ measured in the experiment is converted to the bound for the cross section of dark matter-target atom scattering $\sigma_A$ by the relation 
\begin{eqnarray}
R=N_T n_\chi \int_{E_R} dE_R\ \int_{v} d^3v\  f(v) \frac{\sigma_A m_A}{2 v \mu_A},
\end{eqnarray}
where $N_T$, $n_\chi$, $E_R$, $v$, $m_A$, and $\mu_A$ are the number of target nucleons, the number of dark matter particles, the recoil energy, the velocity of dark matter near the Earth, atomic mass, and reduced mass,  respectively. $f(v)$ represents the distribution function of dark matter velocity in the Solar system.
Thus modification of the distribution function $f(v)$ can affect the bound for dark matter-nucleon interaction in the theoretical models.

Dark matter distribution in the Galaxy is expected roughly spherical, and
we usually adopt Maxwell-Boltzmann distribution 
\[
f_{\mathrm{Max}} (v)=\frac{1}{\left(\pi v_0^2\right)^{3/2}} e^{-(v+v_E)^2/v_0^2}
\]
where $v_0$ is the velocity of the Solar system in the galaxy, and $v_E$ is the Earth's velocity relative to dark matter distribution. Similar isothermal distributions are also derived by gravitational potentials of the Galaxy. 

The question whether such distributions is actually realized, should be answered. 
Halo substructure which can be explained by tidal stream torn off the Sagittarius dwarf galaxy has been reported in the Sloan Digital Sky Survey commissioning data \cite{Ibata:2000ys}. Cosmological N-body simulation with not only dark matter halo but also baryons, also supports the existence of co-rotating dark halo \cite{Ling:2009eh}.
Due to the tidal stream, high velocity tail of the dark matter distribution sharply drops compared to Maxwell distribution. 
Variation from Maxwellian is suggested by debris flow as well, that is similar to the tidal stream but spatially homogenized.

Direct detection with directional sensitivity is next-generation experiment. Measuring both the recoil energy and direction where dark matter particles come from is challenging, and it is more efficient to reject background signals than ordinary measurements. 
Directional sensitivity will also help to clear the dark matter distribution. 
In the paper, we study the possibility to measure the local distribution of dark matter focusing on nuclear emulsion detector. 
Nuclear emulsions have relatively high resolution compared to the gas detectors, therefore suitable to measure the angular distribution of dark matter signals. 
In Section \ref{sec:situation}, supposed scattering in our simulation of directional dark matter search with the nuclear emulsions, is introduced. Numerical results are reviewed in Section \ref{sec:numerical}. 
Conclusions are given in Section \ref{sec:conclusion}.

\section{Dark matter-nucleus scattering in the nuclear emulsion detector}
\label{sec:situation}
In the direct search of dark matter, we detect the recoil energy of dark matter-target atom scattering. 
The nuclear emulsions have several spices for target, mainly, calcium (C), nitrogen (N), oxygen (O), silver (Ag) and bromine (Br). 
Ag and Br are so heavy that they are sensitive to relative to heavy dark matter whose mass is $O(100)$ GeV.
Light atoms, i.e., C, N and O, are suitable to detect light dark matter, though they are subdominant components accounting for $\sim$30\% in weight. 
In the numerical calculation, we simulate the inelastic scattering of dark matter with mass $M_\chi$ to each atom. Mass of the target nucleus is represented as $m_N=0.932 A$ ($N=$C, N, O, Ag, and Br) and A is mass number.

In the nuclear emulsion detector, the recoil energy is related to the track length left by the dark matter-nucleon scattering. Minimal detectable track length is about 100 nm, which corresponds to $\sim$160 keV (Ag and Br), and $\sim$33 keV (C, N and O) for the recoil energy.
We converted the recoil energy to track length  in the nuclear emulsions (See \cite{Nagao:2012gp} for correspondence of the track length and the recoil energy) in the numerical calculation.

In the calculation, z-axis is oriented to the direction of dark matter wind against the Earth, and $\theta$ is the scattering angle between z-axis and the track in the nuclear emulsions. Therefore, $\cos{\theta}=1$ when the atom is scattered to the counter direction of the Solar system movement in the Galaxy.

\section{Numerical Results}
\label{sec:numerical}
Density distributions of the scattering atoms are shown for $M_\chi=$ 200 GeV in Figure 1, 2 and 3, which correspond dark matter distribution of Maxwellian, with tidal stream, and debris flow, respectively. Since the dark matter velocity is relatively suppressed for the case with tidal stream and debris flow, the track length tends to shorter than that for Maxwell distribution. 
Similar figures but for $M_\chi=$ 800 GeV are shown in Figure 4, 5 and 6. 
In the calculation, we suppose realistic composition of the nuclear emulsions, in a word, weight ratio of Ag, Br, C, N, and O are 40\%, 29\%, 12\%, 5\% and 12\%, respectively. Signal tendency for each atom is shown in Appendix, for case of Maxwell distribution. 

\begin{figure}[htbp]
 \begin{minipage}{0.5\hsize}
  \begin{center}
   \includegraphics[width=70mm]{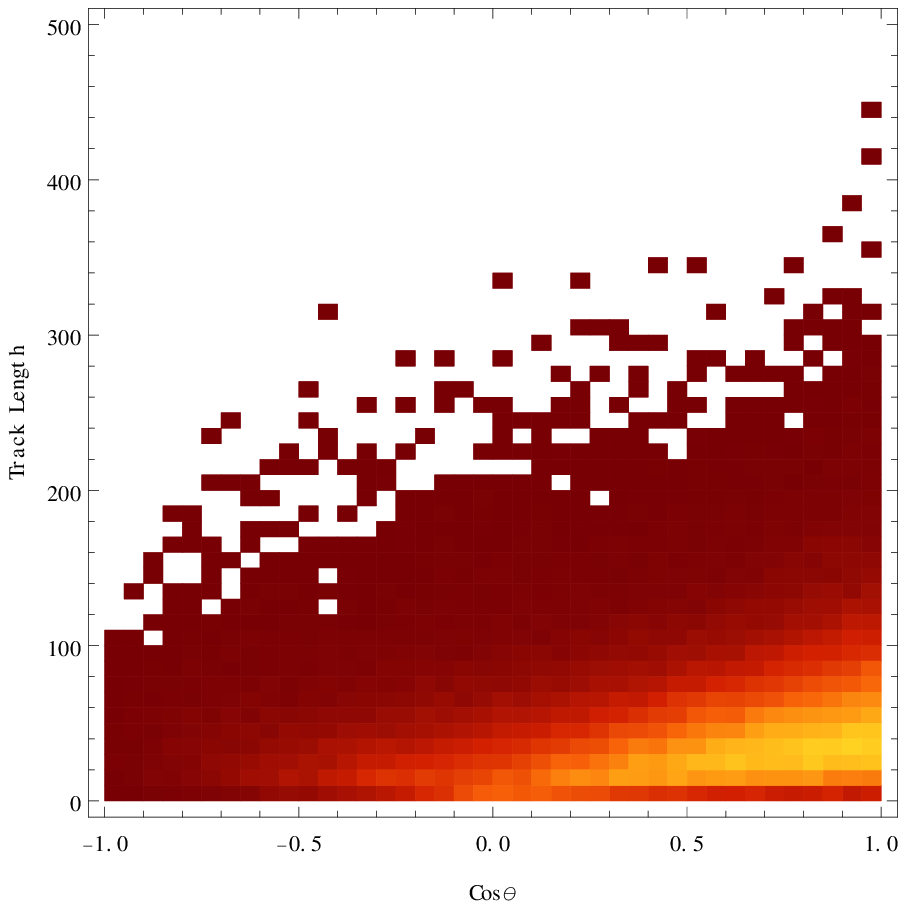}
  \end{center}
  \caption{Angular distribution of track length for Maxwell distribution.}
  \label{fig:one}
 \end{minipage}
\begin{minipage}{0.5\hsize}
  \begin{center}
   \includegraphics[width=70mm]{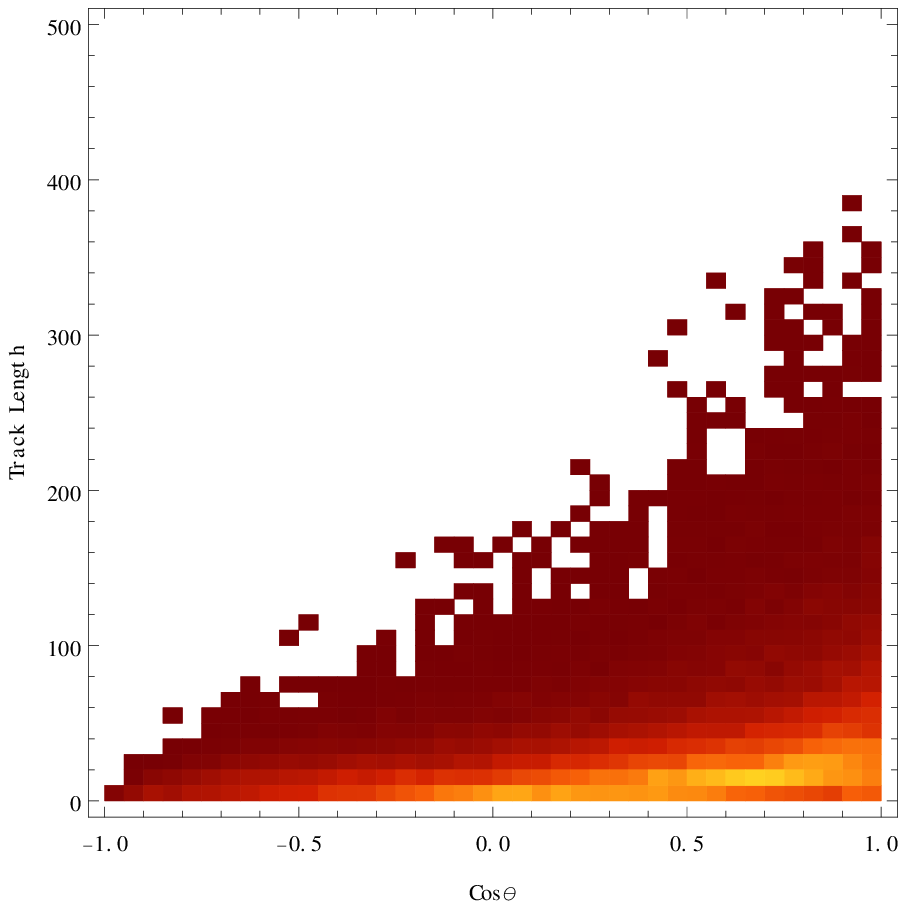}
  \end{center}
  \caption{Angular distribution of track length for the dark matter distribution with tidal stream.}
  \label{fig:one}
 \end{minipage}\\
 \\
 \\
 \vspace{3cm}
  \begin{minipage}{0.5\hsize}
  \begin{center}
   \includegraphics[width=70mm]{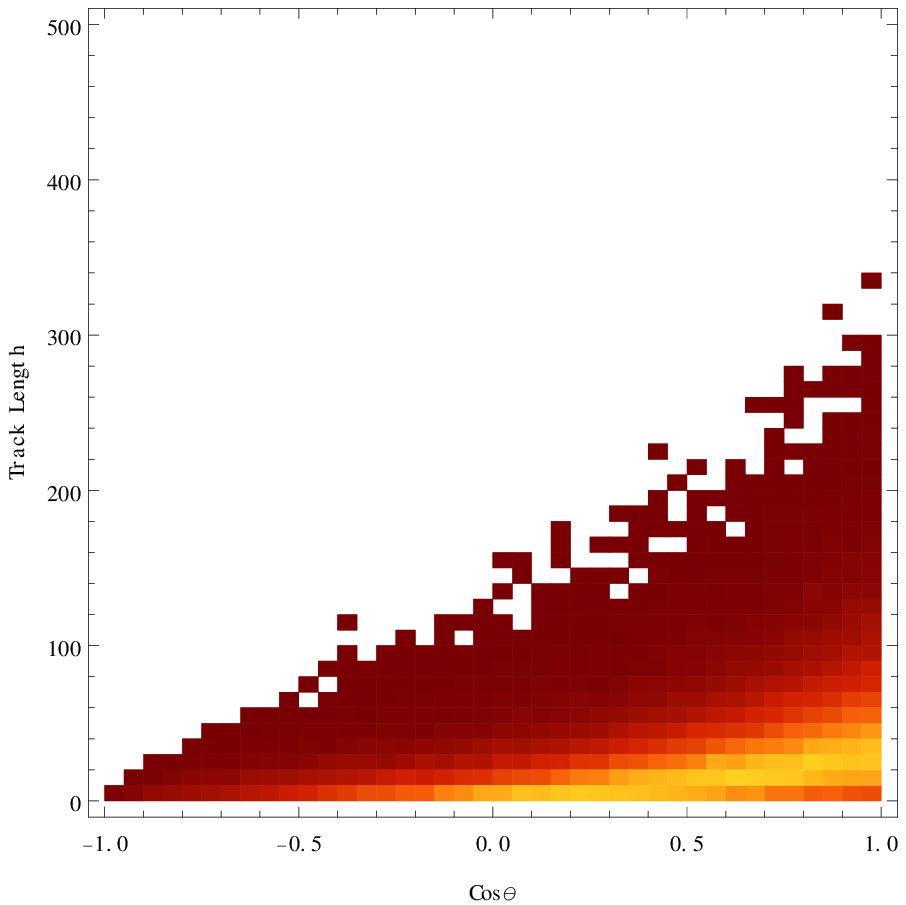}
  \end{center}
  \caption{Angular distribution of track length for the dark matter distribution with debris flow.}
  \label{fig:two}
 \end{minipage}
  \begin{minipage}{0.5\hsize}
 \begin{flushleft}
Density plot of signals of dark matter-nucleon scattering for $M_\chi=200$ GeV. Horizontal and vertical axes represent the scattering angle $\cos{\theta}$ and the track length in the nuclear emulsions in nm unit. Yellow (light gray in printed version) region has more signals than brown (dark) one. Figure 1, 2 and 3 correspond to Maxwellian, the dark matter distributions with tidal stream \cite{Ling:2009eh} and debris flow \cite{Kuhlen:2012fz}.
\end{flushleft}
 \end{minipage}
  \end{figure}
\begin{figure}[htbp]
 \begin{minipage}{0.5\hsize}
  \begin{center}
   \includegraphics[width=70mm]{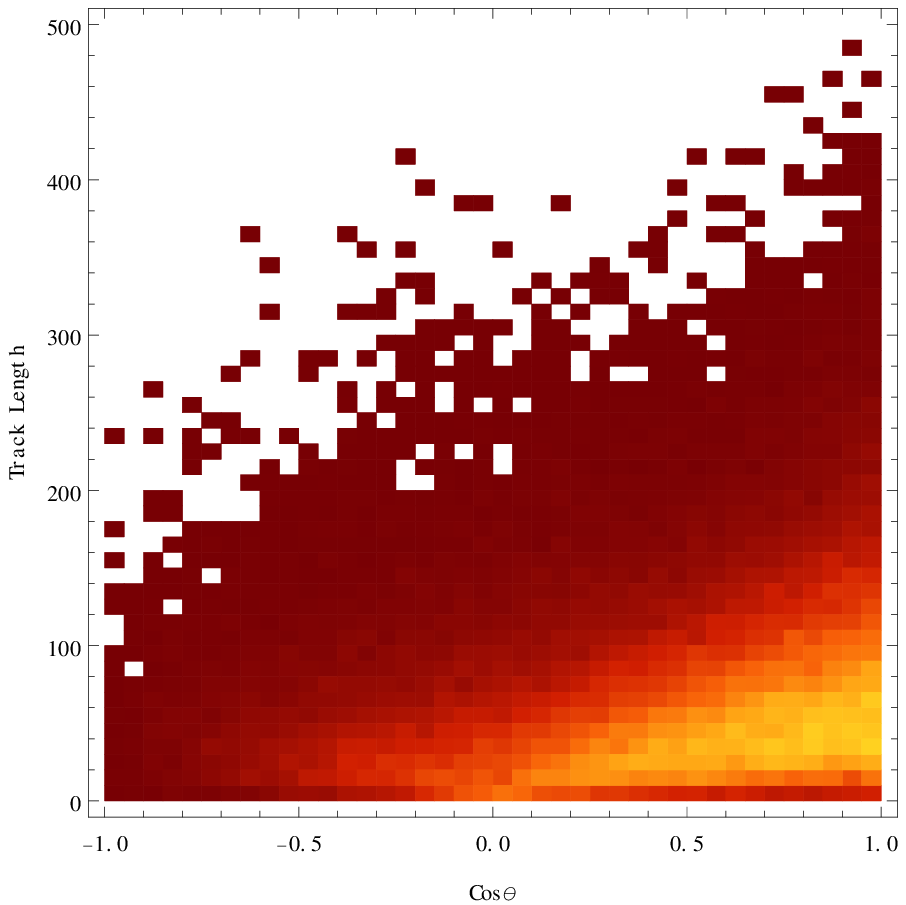}
  \end{center}
  \caption{Angular distribution of track length for Maxwell distribution.}
  \label{fig:one}
 \end{minipage}
 \begin{minipage}{0.5\hsize}
  \begin{center}
   \includegraphics[width=70mm]{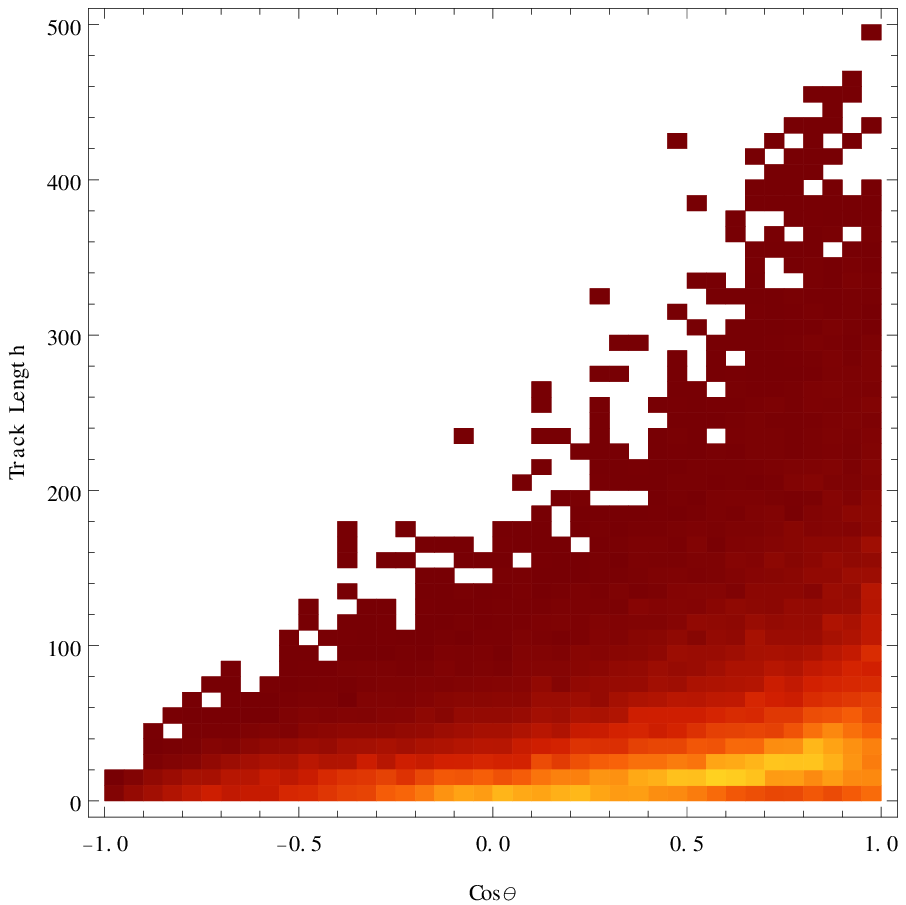}
  \end{center}
  \caption{Angular distribution of track length for the dark matter distribution with tidal stream.}
  \label{fig:two}
 \end{minipage}\\
 \\
 \\
 \vspace{3cm}
  \begin{minipage}{0.5\hsize}
  \begin{center}
   \includegraphics[width=70mm]{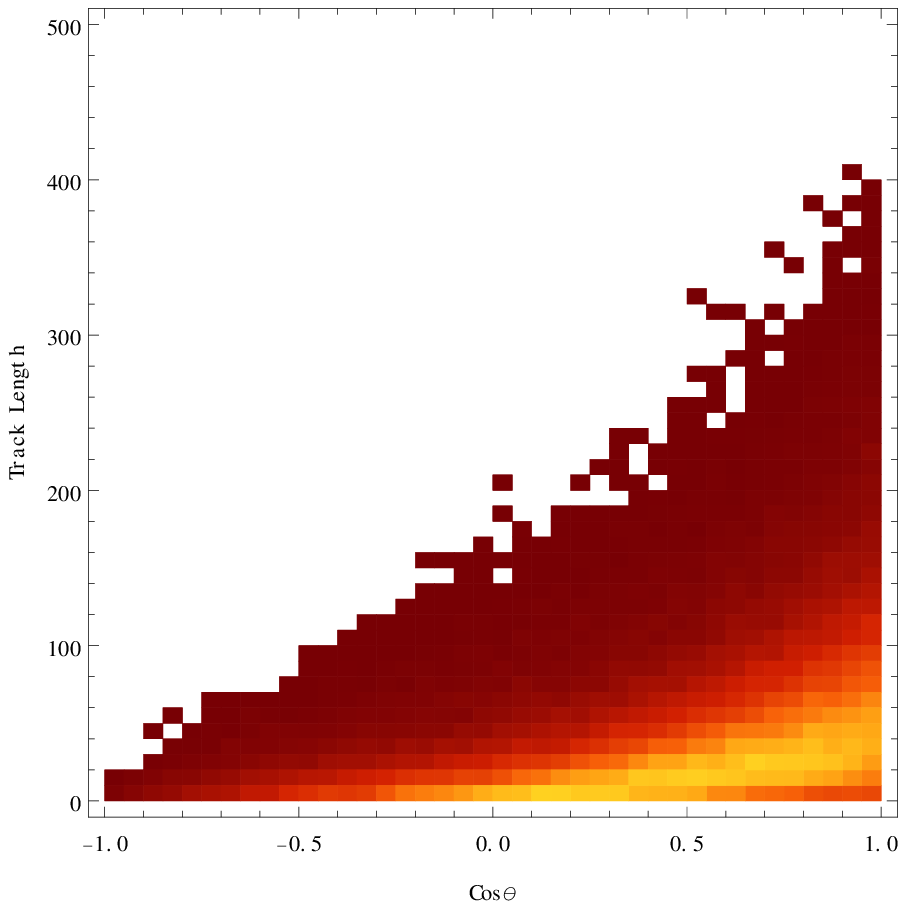}
  \end{center}
  \caption{Angular distribution of track length for the dark matter distribution with debris flow.}
  \label{fig:one}
 \end{minipage}
 \begin{minipage}{0.5\hsize}
 \begin{flushleft}
Legend is same as Figure 1, 2 and 3 except for $M_\chi=800$ GeV. Heavier dark matter than $M_\chi=$ 200 GeV case provides large recoil energy to the scattering atom, and the track length in the nuclear emulsion tends to be long.
\end{flushleft}
 \end{minipage}
\end{figure}

\section{Conclusions}
\label{sec:conclusion}
In this paper, we study the dark matter distribution with direct search of dark matter, focusing on the nuclear emulsion detector. 
Track and angular distributions are shown for standard Maxwell distribution, and the other distribution with tidal stream and debris flow. 
That of tidal stream has suppressed high velocity tail compared to Maxwell distribution, and also dark matter velocity  of  the distribution with debris flow are smaller than that of Maxwell distribution. That makes difference of signal distribution shape even in range where the track is longer than 100 nm, the minimal detectable track length in the nuclear emulsions, if a sufficient number of signals  are provided. Dark matter mass is another unknown quantity which can vary the signal distribution, however, it mainly stretches the track length distribution while keeps the signal distribution shape.

\vspace*{-.3cm}
\section*{Appendix}
Signal distribution for each atom in the nuclear emulsions are shown for Maxwell distribution. We take $M_\chi=$ 80 GeV to demonstrate the difference between heavy and light atoms. Ag and Br atoms are much heavier than the other three, it means it is difficult for them to leave long track compared to light atoms.
\begin{figure}[htbp]
 \begin{minipage}{0.33\hsize}
  \begin{center}
   \includegraphics[width=40mm]{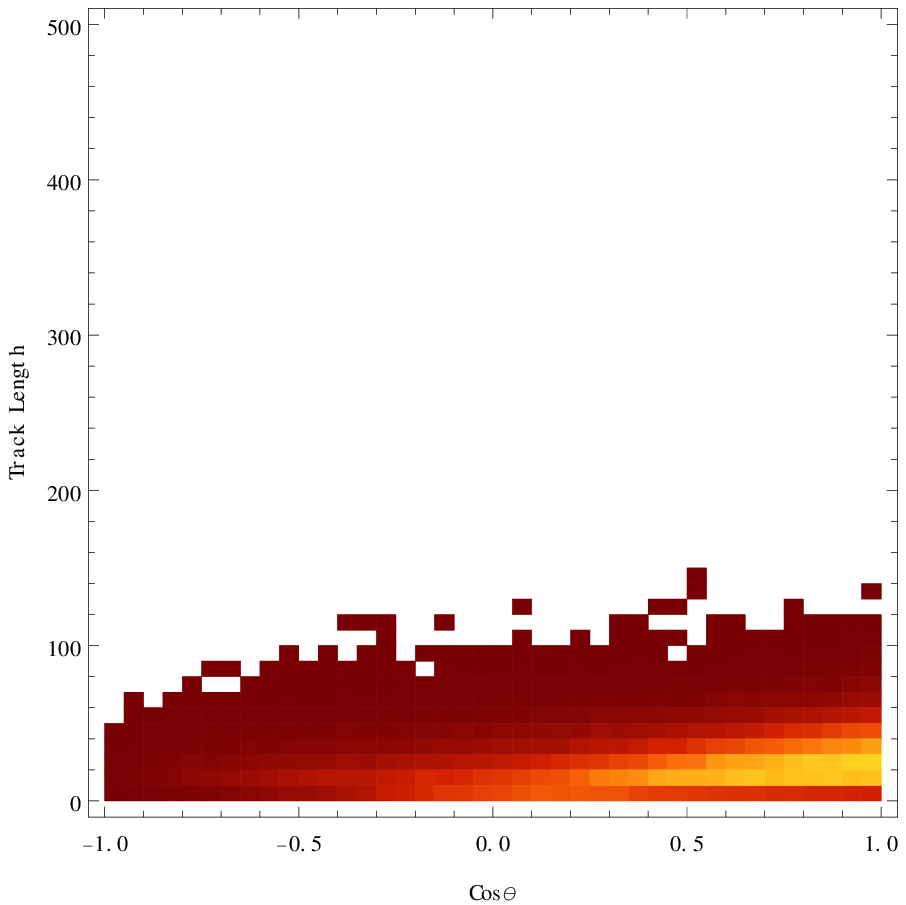}
  \end{center}
  \caption{Case for Ag atom}
  \label{fig:one}
 \end{minipage}
 \begin{minipage}{0.33\hsize}
 \begin{center}
  \includegraphics[width=40mm]{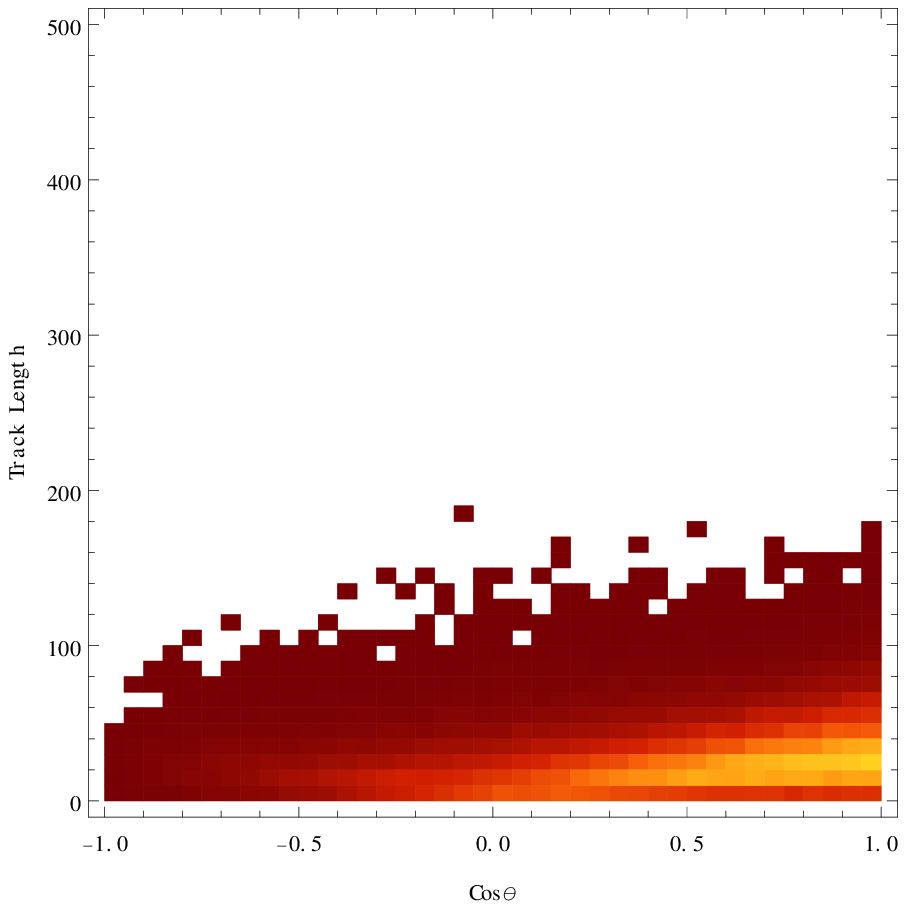}
 \end{center}
  \caption{Case for Br atom}
  \label{fig:two}
 \end{minipage}
 \begin{minipage}{0.33\hsize}
 \begin{center}
  \includegraphics[width=40mm]{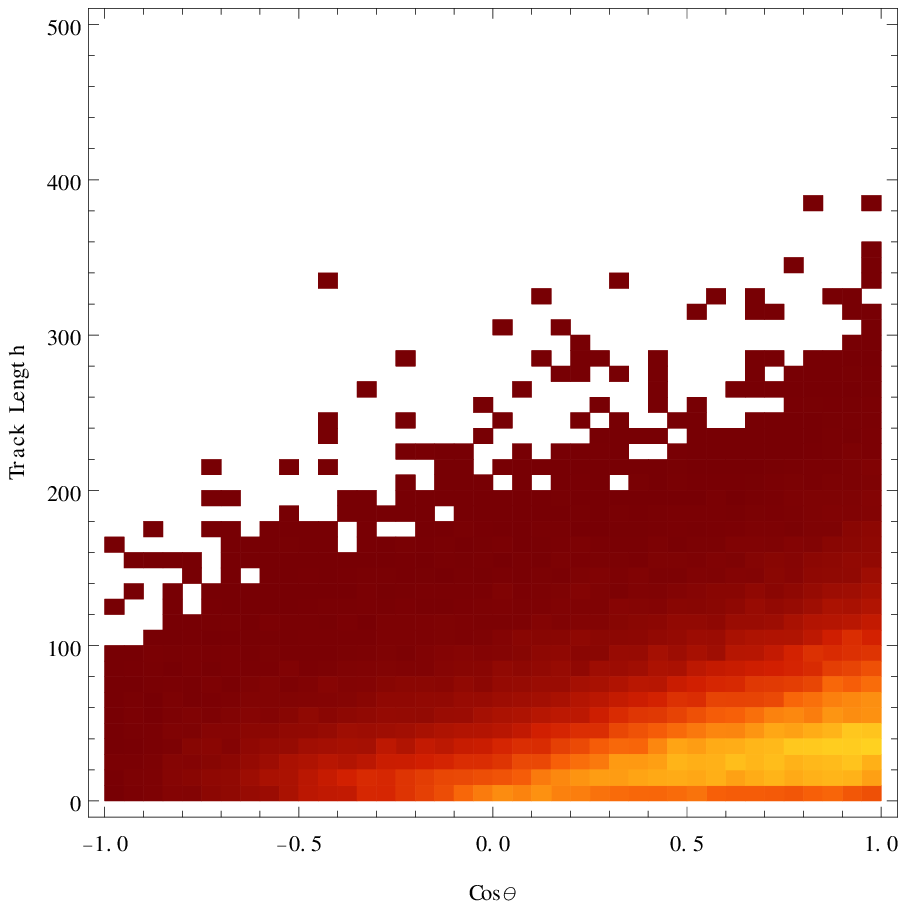}
 \end{center}
  \caption{Case for O atom}
  \label{fig:three}
 \end{minipage}\\ \\ \\ \\
  \begin{minipage}{0.33\hsize}
  \begin{center}
   \includegraphics[width=40mm]{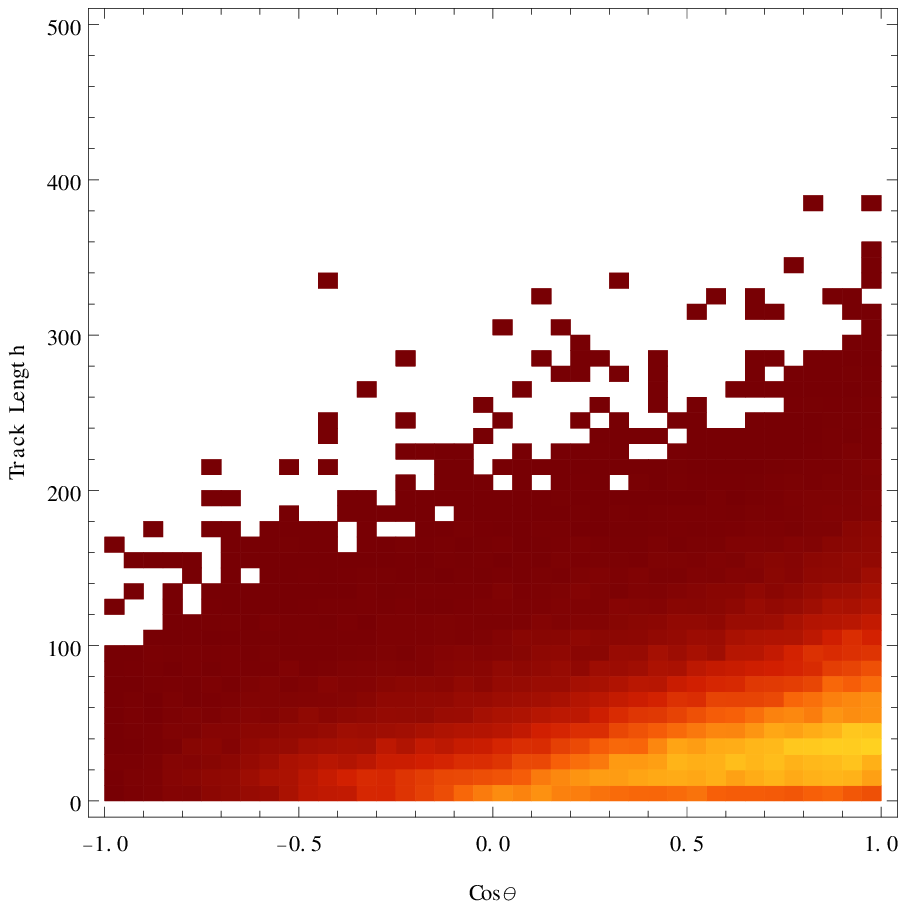}
  \end{center}
  \caption{Case for C atom}
  \label{fig:one}
 \end{minipage}
 \begin{minipage}{0.33\hsize}
 \begin{center}
  \includegraphics[width=40mm]{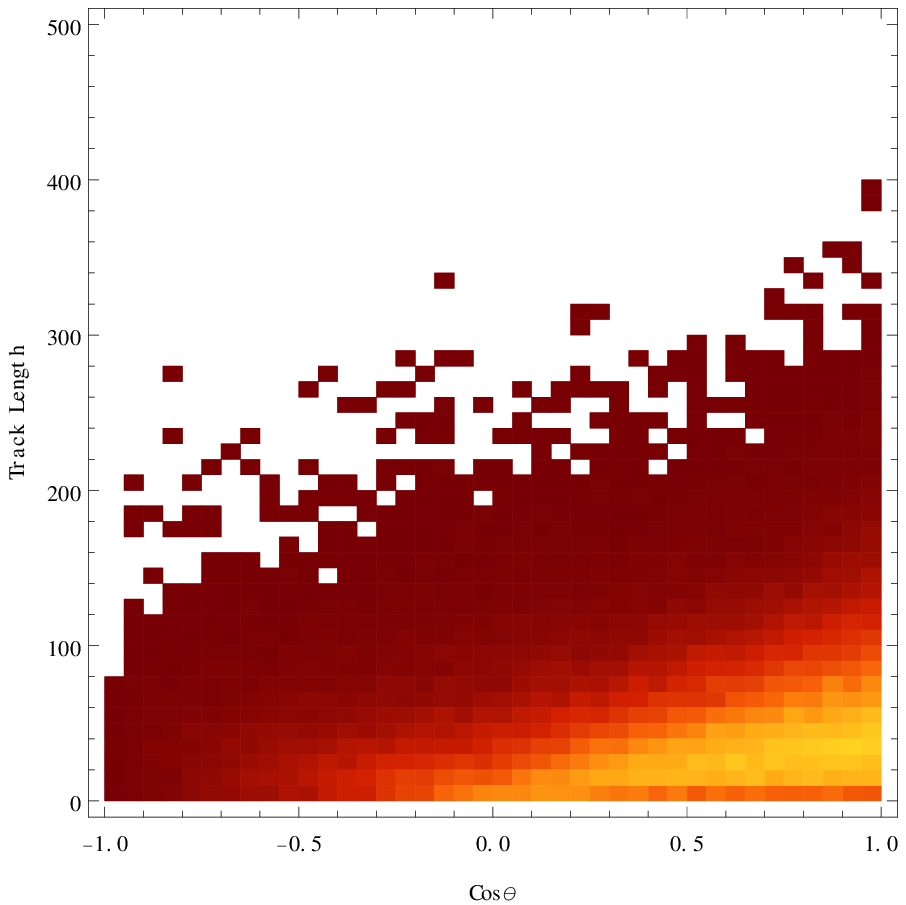}
 \end{center}
  \caption{Case for N atom}
  \label{fig:two}
 \end{minipage}
 \begin{minipage}{0.33\hsize}
 \begin{flushleft}
Legend is same as Figure 1 except for target atoms and $M_\chi=$ 80 GeV. Figure 7-11 corresponds to Ag, Br, C, N and O for target atoms.
 \end{flushleft}
 \end{minipage}
\end{figure}

\end{document}